# CF Recommender system Based on Ontology and Nonnegative Matrix Factorization (NMF)

Sajida Mhammedi[1], Hakim El Massari[1], Noreddine Gherabi[1] and Amnai Mohamed[2]

[1] Lasti Laboratory, National School of Applied Sciences, Sultan Moulay Slimane University, Khouribga, Morocco
[2] LARIT Laboratory, Faculty of Sciences, IbnTofail University, Kenitra, Morocco
`sajida.mhammedi@usms.ac.ma`

**Abstract.** Recommender systems are a kind of data filtering that guides the user to interesting and valuable resources within an extensive dataset. by providing suggestions of products that are expected to match their preferences. However, due to data overloading, recommender systems struggle to handle large volumes of data reliably and accurately before offering suggestions. The main purpose of this work is to address the recommender system's data sparsity and accuracy problems by using the matrix factorization algorithm of collaborative filtering based on the dimensional reduction method and, more precisely, the Nonnegative Matrix Factorization (NMF) combined with ontology. We tested the method and compared the results to other classic methods. The findings showed that the implemented approach efficiently reduces the sparsity of CF suggestions, improves their accuracy, and gives more relevant items as recommendations.

## 1      Introduction

A recommender system is a fully automated system that analyzes user preferences and predicts user behavior. The research interest in this field is still very high, mainly because of the practical significance of the problem. Among the approaches of recommender systems, we distinguish the collaborative filtering approach that relies on the users' ratings to find the most similar ones. CF is believed to be the most effective strategy for dealing with the problem of overloaded data in the context of e-commerce. However, collaborative filtering-based recommender systems typically encounter issues such as data sparsity, scalability, prediction inaccuracy, and recommendation accuracy [1]. [2], for instance, compares recommendation methodologies, whereas [3, 4] classifies recommender systems that use AI algorithms. Furthermore, [4, 5] categorizes the techniques based on suggestion factors. To increase the accuracy of CF recommendations, the researchers construct a semantic-level information model based on an ontological notion to address the issues mentioned above [6, 7]. Furthermore, ontology recommender systems are classified according to the tool, ontology type, and ontology representative language in [3, 4, 8]. Ontologies have been widely used in various fields and contexts because they are incredibly valuable constructs [9, 10]. The work in [11] created a hybrid RS with better semantics by combining the reasoning of ontology-based semantic similarity with conventional elements-based CF. To solve the difficulties of cold start and data sparsity, authors in [12] developed a method that includes the item's semantic domain knowledge as well as the user's social trust network. To enhance the accuracy of a library book recommender, Liao [13]used the Chinese library categorization system as a reference. Ranjbar [14] constructed ANFIS, a multi-standard recommender system, by merging semantic information of items based on ontology with user population statistics. This indicated that incorporating semantic information may increase the prediction accuracy of a multi-standard RS. Tarus [15] introduced a hybrid knowledge-based recommender system that simulates sequential learning patterns and ontologies as user domain and learning course, respectively, to offer online course resources to users. There are a variety of extensive analyses due to the increased research focus on ontology for recommendation systems. For our research proposal, we are interested in hybrid recommender systems, which are slightly more advanced forms of traditional RS but based on CF. Two problems influenced by RS arise the sparsity of the user-element scoring matrix and the lack of recommendation accuracy when the data is significant.

The objective of our work is to remedy these two problems. For this reason, we use the matrix factorization algorithm using one of its methods: the Nonnegative Matrix Factorization (NMF), which is one of the methods used to accelerate the search for content recommendations for users. We also use the conceptualization of items based on ontology. This paper begins by discussing research that is pertinent to the current study. Then moves on to present the background related to our work. The proposed system design is then detailed in-depth. Experiment findings and their evaluation are then presented. Finally, a conclusion was conducted.

## 2 New Hybrid CF Recommender Approach

### 2.1 System Architecture

The suggested system of the CF recommendation algorithm based on semantic modeling of items (ontology) and dimensionality reduction approach using NMF is depicted in Fig. 1. In the beginning, the tree of the hierarchical structure of items was built using ontology conception. The semantic similarity between two concepts was computed utilizing the overlapping connections of the concepts' words. The user-item evaluation matrix was then filled in, and certain missing values in the sparse matrix were estimated using semantic similarity. Furthermore, the semantic similarity threshold was adjusted to improve the content of the matrix data while preserving the source matrix's fundamental attribute characteristics.

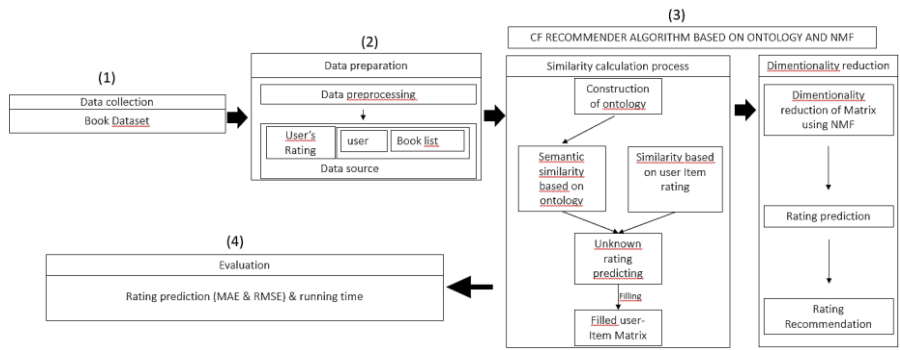

**Fig. 1.** Hybrid System Architecture.

3## 2.2 Similarity calculation process

**Conception of Ontology and Concept Semantic Similarity.**

We should mention that ontologies are employed in various domains and settings since they are extremely useful constructs. As a result, their role differs depending on the framework. This section is inspired by the work done in [16], which suggests an automatic construction of large ontology from a JSON file, according to its domain of use. To use the stated system, we'll use retool utilities to convert our CSV file to a JSON file (https://retool.com/utilities/convert-csv-to-json). We believe that a vocabulary's concepts may be structured into a hierarchical tree, with the most specific terms linked to the more general ones via a parent-child relationship.

The notion of distance between items is based on the similarity of their meaning or semantic content, and semantic similarity is a measure established on a set of documents or words. It's a mathematical method for estimating the strength of a semantic link between language units, concepts, or instances using a numerical description generated by comparing evidence to support or describe their nature. The similarity between two concepts, C1 and C2, is calculated using the ontology's hierarchical structure. Consequently, the resemblance of the two conceptions is comparable to the proximity of two relatives [15]. We assume the following two assumptions:

- The more similar ancestors any two family members share, the closer they are.
- A family element's distance from a common ancestor affects its distance from other family elements.

The number of common ancestors to the genealogy (i.e., the set of concepts from the root to the concept) of two concepts C1 and C2, is the ratio of their similarities[15]. Its definition is as follows:

$$\text{SemSimilarity}\left(\text{Cl}_i, \text{Cl}_j\right) = \frac{2\,\text{deep}\left(\text{nearP}\left(\text{Cl}_i, \text{Cl}_j\right)\right)}{\left[L\left(\text{Cl}_i, \text{nearP}\left(\text{Cl}_i, \text{Cl}_j\right)\right) + L\left(\text{Cl}_j, \text{nearP}\left(\text{Cl}_i, \text{Cl}_j\right)\right) + 2\text{deep}\left(\text{nearP}\left(\text{Cl}_i, \text{Cl}_j\right)\right)\right]}$$

**Predicting Unknown Rating.**
To predict rating, we apply the following equation:

$$\left(\frac{P(U_u)}{R_{U_u}}, b\right) = \frac{\sum_{v \in N(U_u)} \text{SimSimilarity}(U_u, U_v) \times \left(R_{U_v}(b) - R_{U_v}\right)}{\sum_{v \in N(U_u)} \text{SimSimilarity}(U_u, U_v)}$$

where SimSimilarity $(U_u, U_v)$ indicates the similarity between two users. $\left(\frac{P(U_u)}{R_{U_u}}, b\right)$ represents the prediction rating of book b by user u, R denotes the average rating of user u and user v.

## 2.3 Dimensionality Reduction

**Non-Negative Matrix Factorization (NMF).**

For a matrix A of m x n dimensions and non-negative coefficients, i.e., positive or null, NMF can factor it into two non-negative matrices, W and H, having dimensions m x k, k x n respectively. Here, the matrix A is defined as:

$$A_{m \times n} = W_{m \times k} H_{k \times n}$$

Where, A is the original Input Matrix (Linear combination of W & H), W is the Matrix of Features, H refers to coefficient Matrix (Weights associated with W), and k indicates low rank approximation of A (k ≤ min(m, n)) The goal of NMF is dimensionality reduction and feature extraction. Thus, when we define the lower dimension as k, the goal of NMF is to find two matrices W $R^{mk}$ and H $R^{nk}$ having only nonnegative elements. Therefore, using NMF makes it possible to obtain factorized matrices with dimensions significantly smaller than the product matrix. Intuitively, NMF assumes that the original input consists of a set of hidden features represented by each column of the matrix W. Each column of the matrix H represents the "coordinates of a data point" in the matrix W. In simple terms, it contains the weights associated with the matrix W. In this context, each data point represented as a column in A can be approximated by an additive combination of non-negative vectors, defined as columns in W. This technique has since been widely used in many fields with the aim to study the structure of huge sparse matrices.

## 3    Experimental Evaluations Results

### 3.1    Dataset

The BookCrossing (taken from the BookCrossing.com book RS) is an online free book club. This dataset is based on notes rather than tags. Indeed, users are encouraged to evaluate the books they share by assigning a number between 1 and 10, with the higher the score, the better the book's worth. The purchase of books is solely perceived as a sign of contentment, referred to as implicit voting. The dataset we employed for our analysis is openly available and comprises three tables of data: 278858 users who produced 1149780 ratings on 271379 books. We performed a preprocessing phase before utilizing the Bookcrossing dataset, which included the following steps:

- Removing implicit (zero) ratings is the first step. This experiment exclusively looks at explicit evaluations because implicit ratings tend to add noise to the acquired data.
- Eliminate users who have never rated books. Unrated books will never be rated since the suggested system only offers ratings to books that neighbor have rated. And we remove the picture URL field, in the book's data.

### 3.2    implementation and Evaluation

To measure the accuracy of these predictions, the estimated interactions are compared with the actual interactions, i.e., those created by the user. The suggested Hybrid recommender system is evaluated using two precision criteria: MAE and RMSE. Both of these metrics allow for easy interpretation because they are on the same scale as the original scores. We employed a 5-fold cross-validation to calculate the different criteria for each testing subgroup, splitting rating data into five parts and using one as the testing set and the other as the training set at each iteration. Finally, we compared our innovative hybrid strategy's with CF, CB, and CF+NMF approaches. Fig. 3. Ontology-based conceptualizing and matrix factorization have significantly influenced the development of CF recommendation systems. Furthermore, matrix factorization decreased the size of the matrix and led to faster response time, but it also increased the suggestion performance.

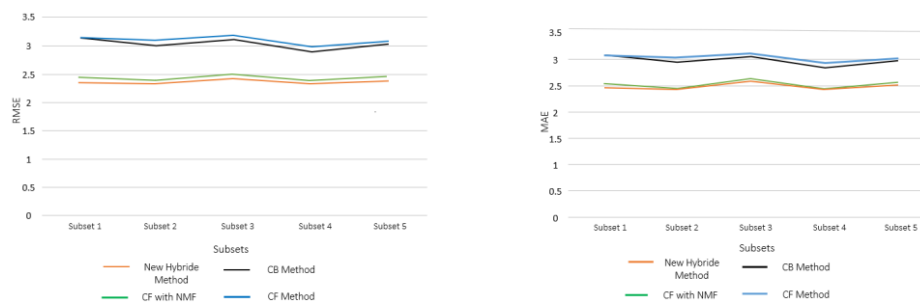

**Fig. 2.** The performances in terms of RMSE and MAE.

## 4      Discussion

Sparsity and scalability are two significant problems when creating RS. As a result, in this work, efforts were made to solve these challenges to enhance the performance of a RS using dimensionality reduction techniques and ontology in the CF. The method is put to the test using real-world data. According to MAE and RMSE measurements, the innovative hybrid CF recommender method effectively improved the performance of CF recommender system. Our findings indicated that hybrid strategies might be used to address the sparsity and scalability issues that face RS. We also found that the technique surpasses previous methods that rely purely on CF, CB, or CF + NMF. Because the recommended technique employs semantic similarity relations for the item in the item-based CF, the suggested approach improves the accuracy of the recommendations. As an out-come, using semantic similarity improves the accuracy of the hybrid approach's suggestions.

## 5      Conclusion and future work

In our work, we have implemented a book RS based on ontology integration and matrix factorization, using (NMF) algorithm. The proposed approach improves the CF system's data sparsity and predictive accuracy. Results proved that ontology-based conceptualization and matrix factorization played an essential role in the RS based on CF. Moreover, the matrix factorization reduced the matrix size, which improved recommendation performance. In the future, we plan to integrate more ML approaches in an attempt to get better results and to improve the performance of recommendations even further.